\documentclass[preprint,showpacs,preprintnumbers,amsmath,amssymb]{revtex4}

\usepackage{graphicx}
\usepackage{dcolumn}
\usepackage{bm}

\begin{document}


\title{Effect of medium dependent binding energies on inferring the temperatures and freeze-out density of disassembling hot nuclear matter from cluster yields}

\author{S. Shlomo, G. R\"opke$^*$, J. B. Natowitz, L. Qin, K. Hagel, 
R. Wada and A. Bonasera$^\dagger$}
\affiliation{
Cyclotron Institute, Texas A\&M University, College Station, 
Texas 77843-3366,\\
$^*$University of Rostock, Institut f\"ur Physik, 18051 Rostock, Germany,\\
$^\dagger$Laboratori Nationali del Sud, INFN, 95123 Catania, Italy}

\date{\today}

\begin{abstract}
We explore the abundance of light clusters in asymmetric nuclear matter at 
subsaturation density. With increasing density, binding energies and wave 
functions are modified due to medium effects. The method of Albergo, Costa, 
Costanzo and Rubbino (ACCR) for determining the temperature and free nucleon 
density of a disassembling hot nuclear source from fragment yields is 
modified to include, in addition to Coulomb effects and flow, also effects 
of medium modifications of cluster properties, which become of importance
when the nuclear matter density is above 10$^{-3}$ fm$^{-3}$. We show how 
the analysis of cluster yields, to infer temperature and nucleon densities, 
is modified if the shifts in binding energies of in medium clusters are 
included. While, at low densities, the temperature calculated from given 
yields changes only modestly if medium effects are taken into account, larger 
discrepancies are observed when the nucleon densities are determined from measured 
yields.
\end{abstract}

\maketitle

 \section{\label{sec:introduction}
 Introduction}

Understanding of nuclear matter at extreme conditions is one of key issues 
to clarify the problems in core-collapse supernovae as well as neutron stars 
and heavy ion collisions. While heavy ion reactions are often employed to 
explore the nuclear matter equation of state (EoS), careful theoretical work 
is needed to analyze the experimental signatures and to reconstruct the 
properties of hot and dense matter from the detected abundances and energy 
distributions of ejectiles or from correlation functions between different 
ejectiles, produced in those reactions. 

Among the interesting observables are the yield ratios of different fragments 
measured in such reactions. In many experiments one commonly observes light 
elements such as neutrons ($n$), protons ($p$), deuterons ($d$), tritons ($t$), $^3$He ($h$), and $^4$He ($\alpha$) (see for example Ref. \cite{kow07} and 
Refs. therein). Larger clusters, typically with mass numbers 
$5 \leq A \leq 20$, are also observed and the production process of these 
fragments must be explained.

The decay of highly excited nuclear matter produced in heavy ion collisions 
is a complex dynamic process which needs, in principle, a sophisticated 
treatment. One simple approach is the freeze-out concept in which the hot and 
dense matter in the initial stage is assumed to reach thermal equilibrium as 
long as reaction rates are high. With decreasing density, the reaction rates 
decrease and the equilibration process becomes suppressed. At that time the 
nuclear thermal and chemical equilibrium is frozen out. Often the description 
of the nuclear matter, in particular the distribution of clusters, is 
calculated within a statistical multifragmentation model assuming nuclear 
statistical equilibrium (NSE) \cite{Bondorf,Gross}. Under the simplifying 
assumption that the final reaction product distribution is identical to the 
cluster distribution at the freeze out point, the thermodynamic parameters 
such as temperature $T$ and particle number densities, $n_n$ and $ n_p$ for 
neutrons and protons, respectively, can be reconstructed from the observed 
abundances. A simple method for extracting the temperature of the fragmenting 
hot system was given by Albergo, Costa, Costanzo and Rubbino (ACCR) 
\cite{albergo}. The method is based on selecting double isotope (or isotone) 
ratios, $R_2$, such that the nucleon chemical potentials are eliminated 
leading to a relation between $R_2$, $T$ and the binding energies of the 
isotopes (isotones). This method has been used in the analysis of a large 
number of experiments. See, for example, the early works of Refs. 
\cite{Poch95, kolom96, Hauger96}.

However, one has to be aware that the dynamic reaction processes do not 
cease abruptly, so that the concept of a unique freeze out time is only 
approximate. A single freeze out time, independent of the species under 
consideration and their dynamical state, the flow and further parameters 
describing non-equilibrium effects, may not exist. In addition, secondary 
(post freeze-out) decay will modify the original distribution which also 
contains excited states, and the decay products will be found in the final 
distribution. We point out that in Ref. \cite{kolom97}, the ACCR method 
was modified to account for the screening due to the Coulomb interactions 
among fragments in the freeze-out volume by using the Wigner-Seitz 
approximation \cite{WS34}. It was found that the corrections for the 
temperature are less than 20\%, though for certain isotone double ratios 
it can be as large as 50\%. In Ref. \cite{shlomo97} the ACCR method was 
modified to account for the effect of radial collective flow. It was found 
that the effect on the extracted temperature is relatively small, but the 
increase in the freeze-out density can be significant for large flow energy. 
It was noted in Ref. \cite{kolom96} that an important improvement of the 
simple ACCR method resulted from taking into account post emission decay 
(secondary decay) processes of particles and, in particular, $\gamma$ 
which modify the freeze-out yield ratios. Without this correction, different 
double ratios $R_2$ associated with selected sets of fragments (different 
thermometers) may result in significantly different temperature $T$. We 
will not discuss these issues further in this work, for a review see for 
example Ref. \cite{shlomo05}.

If the freeze-out density is not very low, i.e. not at baryon number densities 
$n_B \lesssim 10^{-4}$ fm$^{-3}$, the NSE will be modified by medium effects. 
In this paper, the shift of the binding energies of the light elements in hot 
and dense nuclear matter \cite{roe83,roe08} is considered. The changes of the 
mass fractions of different nuclei due to medium effects complicates the 
determination of the temperature $T$ and baryon number density $n_B= n_n+n_p$ 
from yields of the observed products. In the following we show that simple 
NSE approaches like the ACCR thermometer can be improved if in-medium effects 
are taken into account.

\section{The yield thermometer}
Within a quantum statistical approach to nuclear matter \cite{roe83,roe08}, 
using a cluster decomposition for the self-energy we obtain expressions for 
the total proton density
\begin{equation}
\label{EoS:p}
n_p(T, \mu_p, \mu_n) =  \sum_{A,Z} Z n(A,Z)
\end{equation}
and for the total neutron density
\begin{equation}
\label{EoS:n}
n_n(T, \mu_p, \mu_n) =  \sum_{A,Z} (A-Z)  n(A,Z)\,.
\end{equation}
Here, $n(A,Z)$ is the contribution of the $A$-nucleon cluster to the total 
nucleon density. Both equations (\ref{EoS:p}) and (\ref{EoS:n}) may be 
considered as a nuclear matter equation of state (EoS) which determines the 
nucleon densities $n_\tau$ as functions of the temperature and the neutron 
and proton chemical potentials, respectively denoted by $T, \mu_n$ and 
$\mu_p$. Additional thermodynamic quantities such as free energy and other 
thermodynamic potentials are obtained by integration.

Starting with the ideal mixture of different species, where the interaction 
between the species is neglected, the number density of a cluster 
$n^{(0)}(A,Z)$ is given by 
\begin{equation}
\label{yields}
n^{(0)}(A,Z) =  g_{A,Z} \int \frac{d^3 p}{(2 \pi)^3} 
f_{A,Z}[E^{(0)}_{A,Z}(p)],
\end{equation}
with the (Fermi or Bose) distribution function
\begin{equation}
\label{distribution}
f_{A,Z}(E)=
 \frac{1}{e^{\frac{1}{T}[E -Z \mu_p-(A-Z) \mu_n]}-(-1)^A}.
\end{equation}
In Eqs. (\ref{yields}), 
$E^{(0)}_{A,Z}(p) = E^{(0)}_{A,Z} +\hbar^2 p^2/(2 A m)$, where, in the 
non-interacting case considered here, $ E^{(0)}_{A,Z} $ is the ground state 
binding energy, $g_{A,Z}$ is the degeneracy factor of an isolated nucleus 
with mass number $A$ and charge number $Z$, $m$ is the average nucleon mass 
and $\mu_n$ and $\mu_p$ are the chemical potentials of neutrons and protons, 
respectively. Note that in general excited states which are characterized by 
internal quantum numbers may occur, in addition to $\{A,Z\}$. In that case, 
a summation over the excited states, including scattering states, should be 
carried out. The number density is assumed to be proportional to the cluster 
yield, observed after freeze-out. In the non-degenerate limit we have the 
prediction of the cluster yields within NSE models.
\begin{equation}
\label{NSEyields}
Y^{(0)}(A,Z) \propto n^{(0)}(A,Z) = g_{A,Z} \left( \frac{2 \pi \hbar^2}{AmT} 
\right)^{-3/2} e^{-[E^{(0)}_{A,Z} -Z \mu_p-(A-Z) \mu_n]/T}. 
\end{equation}
Inserting $ n^{(0)}(A,Z) $ for the cluster densities $n(A,Z)$, 
Eqs. (\ref{EoS:p}) and (\ref{EoS:n}) read 
\begin{eqnarray}
\label{EoS:p0}
n_p^{(0)}(T, \mu_p, \mu_n)& =&  \sum_{A,Z} Z n^{(0)}(A,Z)\,,
\nonumber\\
n_n^{(0)}(T, \mu_p, \mu_n) & = &  \sum_{A,Z} (A-Z)  n^{(0)}(A,Z)\,,
\end{eqnarray}
for the total proton and neutron densities, respectively. They are 
approximations to the nuclear matter EOS, reflecting NSE.

Let us now consider the observed cluster yields $Y(A,Z)$ which are 
proportional to the number density fractions $n(A,Z)/n_B$ of the cluster 
$\{A,Z\}$. We introduce the (single) ratio of the observed cluster 
yields
\begin{equation}
\label{singleratio}
 R_{(AZ),(A'Z')} = \frac{Y(A,Z)}{Y(A',Z')}.
\end{equation}
If we accept the concept of NSE, identifying the observed cluster yields with 
the predicted ones, Eq. (\ref{NSEyields}), we can get an estimation for the 
temperature $T $ and the chemical potentials $\mu_n, \mu_p  $ of nuclear 
matter produced in heavy ion collisions in a fashion similar to that employing 
the well-known Saha equation in plasma physics \cite{Saha}. Specifically, 
since the abundances of different bound states are determined by the 
temperature and the chemical potentials, observed yield ratios can be used to 
determine these parameters. A simple method to derive the temperature of the 
hot system was given by ACCR \cite{albergo}, assuming NSE (\ref{NSEyields}) 
and selecting double isotope ratios such that the nucleon chemical potentials 
are eliminated. In particular, the H-He thermometer considers the double ratio 
$R^{(0)}_{\rm HHe}$ of cluster yields $Y^{(0)}$,
\begin{equation}
\label{albergo}
R^{(0)}_{\rm HHe}=\frac{Y^{(0)}(^2{\rm H})\,Y^{(0)}(^4{\rm He})}{Y^{(0)}
(^3{\rm H})\,Y^{(0)}(^3{\rm He})}=\frac{3\cdot 1}{2 \cdot 2} 
\left(\frac{2\cdot 4}{3 \cdot 3} \right)^{3/2}e^{-[E^{(0)}_{^2{\rm H}}
+E^{(0)}_{^4{\rm He}}-E^{(0)}_{^3{\rm H}}-E^{(0)}_{^3{\rm He}}]/T},
\end{equation}
where the degeneracy and mass factors are explicitly included. Identifying 
the double ratio
\begin{equation}
\label{doubleratio}
R_{\rm HHe}=\frac{ R_{(^2{\rm H}),(^3{\rm H})}}{ R_{(^3{\rm He}),(^4{\rm He})}}
\end{equation}
of the observed cluster yields for $d\, (^2{\rm H}),\,t\, (^3{\rm H}),\,h\, 
(^3{\rm He})$ and $\alpha \,(^4{\rm He})$ with the prediction according to 
the NSE, $R_{\rm HHe}=R^{(0)}_{\rm HHe}$, we deduce the ACCR temperature 
$T^{(a)}_{\rm HHe} $ ($=T^{(0)}_{\rm HHe} $) corresponding to the observed 
double ratio $R_{\rm HHe}$ as 
\begin{equation}
\label{albergoT0}
 T^{(a)}_{\rm HHe}= \frac{14.325 \,\rm MeV}{\ln [1.591\, R_{\rm HHe}]}\,.
\end{equation}
The constants $14.325 \, {\rm MeV} = -[-2.225-28.3+8.482+7.718]$ MeV and 
$1.591 = 9/ \sqrt{32}$ reflect the ground states binding energies, spins 
and mass numbers of the ejectiles as given in Eq. (\ref{albergo}). 

Other combinations of isotopes can be used to construct double ratios of 
cluster yields  where, within a simple NSE, the chemical potentials cancel 
out so that an ACCR temperature can be derived directly. Thus, thermometers 
based on the yields of other nuclei such as lithium or beryllium isotopes 
can be introduced. Similar approaches are used in hadron production to 
derive the temperature for the quark-gluon plasma phase transition 
\cite{BraunMunzinger}, or in plasma physics \cite{Saha,ChenHan} considering 
spectral line intensities of different ionization states of radiating atoms.

The advantage of the double ratio is that, within NSE, it does not contain 
the density, because the chemical potentials cancel. Therefore the 
temperature determination seems to be insensitive with respect to the 
determination of other parameters. These other parameters, in particular the 
chemical potentials, are observed if, in addition to the double ratios like 
$R_{\rm HHe}$, the single ratios $ R_{(AZ),(A'Z')} = Y(A,Z)/Y(A',Z')$ of 
yields are considered.

There are some objections to inferring the parameter values of hot dense 
matter from the cluster yields. First, we have to take into account that 
collisions lead to initially inhomogeneous system evolving in time. Even 
assuming local thermal equilibrium, one has to separate the ejectiles 
arising from different sources. In efforts to do this, the  H-He 
thermometer has been applied to the double ratio $R_{v_{\rm surf}}$ of 
cluster yields $Y(A,Z)$ for clusters with the same surface velocity 
\cite{kow07}. In that case an additional factor $ \sqrt{(9/8)} $ arises in 
the temperature equation when the number densities as a function of velocity 
are employed.

An important improvement of the simple NSE model was to take into account 
secondary decay processes which modify the freeze-out yield ratios. This 
has been considered in different papers. This correction is essential to 
reduce the differences of the ACCR temperatures obtained from different 
thermometers \cite{kolom96}.  

The simple NSE is based on a chemical picture considering a non-interacting, 
ideal mixture of different components, which is in chemical equilibrium due 
to reactive collisions as described by the mass action law. Such an approach 
is valid in the low-density limit, and related expressions such as virial 
expansions can be taken as a benchmark in that limit \cite{hor06}. With 
increasing density, modifications arise which are based on taking the 
interactions between the different components into account. Thus, as the 
density increases, corrections to the ACCR approach to derive the temperatures 
of hot and dense matter are expected. In earlier work, the effects of the 
screening of the Coulomb interaction and of the flow on the freeze-out 
density and temperature of disassembling hot nuclei have been considered 
\cite{shlomo05,shlomo97}. The effect of screening of the Coulomb interaction 
becomes of importance for heavy nuclei at densities near to the saturation 
density.

The main topic we address in this paper is the required modification of the 
description of the matter as an ideal, noninteracting mixture of different 
components when densities are not low enough to justify this assumption.  
Despite the fact that the nucleon-nucleon interaction is short-ranged, the 
interaction between the free nucleons as well as nucleons bound in clusters 
is negligible only below about $10^{-3}$ times the nuclear saturation 
density, i.e. at  baryonic densities $n_B \lesssim 10^{-4}$ fm$^{-3}$. An 
important question is the role of medium effects due to the nucleon-nucleon 
interactions. In fact, our work indicates that the concept of the simple NSE 
considering hot and dense nuclear matter as an ideal mixture of different 
clusters is not appropriate to describe disassembling hot matter at densities 
at and above approximately one tenth of saturation density. We address this 
in the following section.

\section{Medium modification of cluster properties}

Recent progress in the description of clusters in low density nuclear matter 
\cite{roe05,roe06,sumi08,roe08} enables us to evaluate the abundance of 
deuterons, tritons and helium nuclei in a microscopic approach, taking the 
influence of the medium into account. Within a quantum statistical approach 
to the many-particle system, we determine the single-particle spectral 
function, which allows calculation of the density of the nucleons as a 
function of $T, \mu_n$ and $\mu_p$. The main ingredient is the self-energy 
$\Sigma (1,z)$ which is treated in different approximations. The 
single-particle spectral function contains the single-nucleon quasiparticle 
contribution, $E^{\rm qu}(1) =E^{\rm qu}_{1,Z}(p)$ or $E^{\rm qu}_{\tau}(p)$, 
where $\tau $ denotes isospin (neutron or proton). The quasiparticle energy 
follows from the self-consistent solution of 
$E^{\rm qu}_{\tau}(p) = \hbar^2 p^2/(2 m_\tau) + {\rm Re} 
\Sigma[p,E^{\rm qu}_{\tau}(p)]$. 

Expressions for the  single-nucleon quasiparticle energy 
$E^{\rm qu}_{\tau}(p)$ can be given by the Skyrme mean-field parametrization 
\cite{vau72} or by more sophisticated approaches such as relativistic mean 
field approaches \cite{typ05} and relativistic Dirac-Brueckner Hartree Fock 
\cite{mar07} calculations. In the effective mass approximation, the 
single-nucleon quasiparticle dispersion relation reads
\begin{equation}
E_{\tau}^{\rm qu}(p) = \Delta E^{\rm SE}_{\tau}(0) 
+\frac{\hbar^2}{2 m_\tau^*}p^2 + {\mathcal O}(p^4)\,,
\end{equation} 
\label{quasinucleonshift}
where the quasiparticle energies are shifted by $\Delta E^{\rm SE}_{\tau}(0)$, 
and $m_\tau^*$ denotes the effective mass of  neutrons ($\tau=n$) or protons 
($\tau=p$). Both quantities,  $\Delta E^{\rm SE}_{\tau}(0)$ and $m_\tau^*$, 
are functions  of $T, n_p$ and $n_n$ characterizing the surrounding matter. 
Empirical values for the effective mass near the saturation density are 
different from  the nucleon mass. In the low-density region considered here, 
the effective mass may be replaced by the free nucleon mass. For calculating 
the yields, the quasiparticle shift $\Delta E^{\rm SE}_{\tau}(0)$ can be 
implemented in a renormalization of the corresponding chemical potentials.

In addition to the $\delta$-like quasiparticle contribution, the contribution 
of the bound and scattering states can also be included in the single-nucleon 
spectral function, by analyzing the imaginary part of $\Sigma (1,z)$. Within 
a cluster decomposition, $A$-nucleon T matrices appear in a many-particle 
approach. These T matrices describe the propagation of the $A$-nucleon 
cluster in nuclear matter. In this way, bound states contribute to the EoS,  
$n_\tau = n_\tau(T,\mu_n, \mu_p)$, see Refs. \cite{roe83,sch90}. In the 
low-density limit, the propagation of the $A$-nucleon cluster is determined 
by the energy eigenvalues of the corresponding nucleus, and the simple EoS, 
(\ref{EoS:p}) and (\ref{EoS:n}), results.

For the nuclei embedded in nuclear matter, an effective wave equation can 
be derived \cite{roe83,roe08}. The $A$-particle wave function and the 
corresponding eigenvalues follow from solving the in-medium Schr\"odinger 
equation  
\begin{eqnarray}
&&[E^{\rm qu}(1)+\dots + E^{\rm qu}(A) - E^{\rm qu}_{A \nu}(p)]
\psi_{A \nu p}(1\dots A)\nonumber \\ &&
+\sum_{1'\dots A'}\sum_{i<j}[1-\tilde f(i)- \tilde f(j)]V(ij,i'j')
\prod_{k \neq   i,j} \delta_{kk'}\psi_{A \nu p}(1'\dots A')=0\,.
\label{waveA}
\end{eqnarray}
This equation contains the effects of the medium in the single-nucleon 
quasiparticle shifts as well as in the Pauli blocking terms. 

The in medium Fermi distribution function $\tilde f(1)=\{\exp[E^{\rm
qu}(1)/T-\tilde \mu_1/T] +1 \}^{-1}$ contains the effective chemical 
potential $\tilde \mu_1$ which is determined by the total proton or neutron 
density, calculated in the quasiparticle approximation, 
$n_\tau = \Omega^{-1} \sum_1 \tilde f(1) \delta_{\tau_1,\tau}$. It describes 
the occupation of the phase space neglecting any correlations in the medium. 
In the low-density and non-degenerate limit ($\tilde \mu_\tau <0  $), 
assuming the effective mass approximation for the nucleon quasiparticle 
dispersion relation, we eliminate $\tilde \mu_\tau  $ using
\begin{equation}
\label{f1}
\tilde f_\tau(p) = \frac{1}{\exp[E_\tau^{\rm qu}(p)/T- \tilde \mu_\tau/T] +1} 
\approx \frac{n_\tau}{2} \left(\frac{2 \pi \hbar^2}{m^*_\tau T}\right)^{3/2} 
e^{-\frac{\hbar^2 p^2}{2 m^*_\tau T}}\,.
\end{equation}

The solution of the in-medium Schr\"odinger equation (\ref{waveA}) can be 
obtained in the low density region by perturbation theory. In particular, 
the quasiparticle energy of the $A$-nucleon cluster follows as
\begin{equation}
E^{\rm qu}_{A,Z}(p)= E_{A,Z}^{(0)}+\frac{\hbar^2p^2}{2 A m}+
\Delta E_{A,Z}^{\rm SE}(p)+\Delta E_{A,Z}^{\rm Pauli}(p)\,.
\label{finalshift}
\end{equation}
Additional contributions such as the Coulomb shift 
$\Delta E_{A,Z}^{\rm Coul}(p)$, which can be evaluated for dense matter in 
the Wigner-Seitz approximation \cite{kolom97,WS34,shlomo05,roe84}, will not 
be considered here since the values of $Z$ are small and the densities are 
low. The general formalism also allows us to describe pairing or quartetting, 
but this will not be done here. Disregarding the effects due to the change of 
the effective mass, the self-energy contribution to the quasiparticle shift 
is determined by the contribution of the single-nucleon shift
\begin{equation}
\label{delArigid}
\Delta E_{A,Z}^{\rm SE}(0)= (A-Z) \Delta E_n^{\rm SE}(0)+ 
Z \Delta E_p^{\rm SE}(0)\,.
\end{equation}
Inserting the medium-dependent quasiparticle energies in the distribution 
functions $f_{A,Z}[E^{\rm qu}_{A,\nu}(p)]$, Eq. (\ref{distribution}), this 
contribution to the quasiparticle shift can be included by renormalizing the 
chemical potentials $\mu_n$ and $\mu_p$. 

The most important effect on the calculation of the yields of light elements 
comes from the Pauli blocking terms in Eq. (\ref{waveA}) in connection with 
the interaction potential. This contribution is restricted only to the bound 
states so that it may lead to the dissolution of the nuclei if the density 
of nuclear matter increases. The corresponding shift 
$\Delta E_{A,Z}^{\rm Pauli}(p)$ can be evaluated in perturbation theory 
provided that the interaction potential and the ground state wave function 
are known. After angular averaging, the Pauli blocking shift can be 
approximated as
\begin{equation}
\label{delpauli0P}
\Delta E_{A,Z}^{\rm Pauli}(p) \approx \Delta E_{A,Z}^{\rm Pauli}(0) \, 
e^{-\frac{\hbar^2 p^2}{2 A^2 m T}}\,.
\end{equation}
The shift of the binding energy of light clusters at zero total momentum 
which is of first order in density  \cite{roe05,roe06} has been calculated 
recently \cite{roe08}. Besides neutrons ($n$) and protons ($p$), light 
elements deuterons $^2$H, $\{A,Z\}=d$, tritons $^3$H, 
$\{A,Z\}=t$, hellions  $^3$He, $\{A,Z\}=h$, and $\alpha$-particles  $^4$He, 
$\{A,Z\}=\alpha$ have been considered. The interaction potential and the 
nucleonic wave function of the few-nucleon system have been fitted to the 
binding energies and the root mean square (rms) radii of the corresponding 
nuclei. The following results (in MeV, fm) are obtained for the binding 
energy shifts.
\begin{eqnarray}
\label{shifts}
 \Delta E_d^{\rm Pauli}& =&\left\{ \frac{38384}{(1+ \frac{22.52}{T})^{1/2}}
-0.39402 \, e^{0.049418 \left(1+\frac{22.52}{T} \right)} {\rm Erfc} 
\left[0.2223 \left(1+\frac{22.52}{T}\right)^{1/2}\right] \right\} 
\frac{n_p + n_n}{T^{3/2}}\,,\nonumber \\
 \Delta E_t^{\rm Pauli} & = &  3389.7 \,[1+0.13347 \,T]^{-3/2}
\left(\frac{2}{3}n_p +\frac{4}{3} n_n\right)\,, \nonumber \\
 \Delta E_h^{\rm Pauli} & = &  3901.5 \,[1+0.16455 \,T]^{-3/2} 
\left(\frac{4}{3} n_p +\frac{2}{3} n_n\right)\,, \nonumber \\
 \Delta E_\alpha^{\rm Pauli} & = &  4716.0 \,[1+0.09372 \,T]^{-3/2} 
\left(n_p + n_n\right)\,.
\end{eqnarray}
These results describe only the linear shifts as functions of the nucleon 
densities. The differences between the values for $ \Delta E_t^{\rm Pauli} 
$ and $ \Delta E_h^{\rm Pauli} $ are mainly caused by different values 
of the rms radii for these two nuclei. With increasing density, higher orders 
terms with respect to the densities also become relevant.

It can be shown \cite{roe84} that the EoS can be evaluated as in the 
non-interacting case (\ref{yields}) given above, except that the number 
densities of clusters must be calculated with the quasiparticle energies,
\begin{equation}
\label{quyields}
n^{\rm qu}(A,Z) =  g_{A,Z} \int \frac{d^3 p}{(2 \pi)^3} f_{A,Z}
[E^{\rm qu}_{A,Z}(p)]\,.
\end{equation}
In the cluster-quasiparticle approximation, the EoS, (\ref{EoS:p}) and 
(\ref{EoS:n}), reads 
\begin{eqnarray}
\label{EoS:pqu}
n_p^{\rm qu}(T, \mu_p, \mu_n) & = &  \sum_{A,Z} Z n^{\rm qu}(A,Z)\,,\nonumber\\
n_n^{\rm qu}(T, \mu_p, \mu_n) & = &  \sum_{A,Z} (A-Z)  n^{\rm qu}(A,Z)\,,
\end{eqnarray}
for the total proton and neutron density, respectively.

This result is an improvement of the NSE and allows for the smooth transition 
from the low-density limit up to the region of saturation density. The bound 
state contributions to the EoS fade with increasing density because they 
merge with the continuum of scattering states. This improved NSE, however, 
does not contain the contribution of scattering states, in particular 
resonances appearing in the continuum of scattering states when bound states 
merge with the continuum. For the treatment of scattering states in the 
two-nucleon case, as well as the evaluation of the second virial coefficient, 
see Refs. \cite{hor06,sch90}. We will also not consider the formation of heavy 
elements here. This limits the present results to the range of parameters 
$T, n_n$ and $ n_p$, where the EoS is determined only by the light elements. 
For a more general approach to the EOS which takes also the contribution of 
heavier clusters into account see Ref. \cite{roe84}.

\section{Improved thermometer and density determination including medium 
effects}

To show the effect of in-medium corrections, we start with a temperature 
$T$ and densities $n_n$ and $ n_p$ and calculate the corresponding yields  
$Y^{\rm qu}(A,Z) $, taking the in-medium shifts into account. Then we use 
these yields to infer the parameter values $T^{(a)} (=T^{(0)}) , 
n^{(a)}_n (=n^{(0)}_n)$ and $n^{(a)}_p (=n^{(0)}_p)$, using the ACCR 
relations which were derived neglecting in-medium corrections. In this way 
we obtain for given ratios of cluster yields $R_{(A,Z),(A',Z')}$, 
Eq. (\ref{singleratio}),  $\{T, n_n, n_p\}$ which we identify as the values 
which would be derived from experiments if the medium effects are considered 
and those $\{T^{(a)}, n^{(a)}_n, n^{(a)}_p\}$, derived without taking the 
medium effects into account. Comparing these sets of the parameters we 
demonstrate how the medium modification of the binding energy of light nuclei, 
Eq. (\ref{quyields}) can modify the results determined from the experimental 
yields of light clusters. The three ratios necessary to determine three 
thermodynamic parameters are derived here from the four yields of the light 
clusters $Y(^2{\rm H}),\,Y(^3{\rm H}),\,Y(^3{\rm He})$ and $Y(^4{\rm He})$.

We first compare results of the determination of the temperature  $T$ from 
cluster yields, if the in-medium quasiparticle shifts are taken into account, 
with the temperature $T^{(a)}$ determined from the same yields if medium 
effects are neglected. In particular, the temperature $T_{\rm HHe}$ is not 
related in a simple way to the double ratio $R_{\rm HHe}$, Eq. 
(\ref{doubleratio}). Considering the yields in quasiparticle approximation, 
we have
\begin{equation}
R_{\rm HHe}=\frac{Y^{\rm qu}(^2{\rm H})\,Y^{\rm qu}(^4{\rm He})}
{Y^{\rm qu}(^3{\rm H})\,Y^{\rm qu}(^3{\rm He})}. 
\end{equation}
If we take the yields $Y^{\rm qu}(A,Z) \propto n^{\rm qu}(A,Z)$, 
Eq. (\ref{quyields}), in the non-degenerate case, we obtain the relation
\begin{equation}
\label{albergoTqu}
 T_{\rm HHe}= \frac{1}{\ln [1.591\, R_{\rm HHe}]} \left( 14.325 \,\rm MeV + 
\Delta E_d^{\rm Pauli}  + \Delta E_\alpha^{\rm Pauli} - 
\Delta E_t^{\rm Pauli} -  \Delta E_h^{\rm Pauli} \right)\,,
 \end{equation}
The energy shifts are functions of temperature and densities so 
that this relation has to be solved self-consistently.

On the other hand, neglecting in-medium corrections, we find from the same 
double ratio the apparent ACCR temperature $T^{(a)}_{\rm HHe}$ according to 
Eq. (\ref{albergoT0}). Using, in the non-degenerate case, Eq. 
(\ref{NSEyields}), the relation between both quantities is given by
\begin{equation}
\label{T0T}
 T^{(a)}_{\rm HHe}(T,n_n, n_p) = \left[ 1+\frac{ \Delta E_t^{\rm Pauli} +  
\Delta E_h^{\rm Pauli} - 
\Delta E_d^{\rm Pauli}  - \Delta E_\alpha^{\rm Pauli} }{ E_t^{(0)} +  
E_h^{(0)} - 
E_d^{(0)}  - E_\alpha^{(0)} }\right]^{-1} T_{\rm HHe}\,.
\end{equation}
In the approximations considered here The self-energy contributions to the 
shifts disappear, in a fashion similar to the chemical potentials. In Figure 
1 we show the ratio between the ACCR temperature $T^{(a)}$ and $T$ as a 
function of the baryon density $n_B = n_p + n_n$ for various values of $T$.

Similarly, the densities can be estimated by considering single ratios 
$R_{(A,Z),(A',Z')}$, Eq. (\ref{singleratio}). If the shifts of the binding 
energies due to medium effects are neglected, we have
\begin{equation}
 R^{(a)}_{(A,Z),(A',Z')}=\frac{g_{A,Z} A^{3/2}}{g_{A',Z'} {A'}^{3/2}}
e^{-[E^{(0)}_{A,Z}-E^{(0)}_{A',Z'}-(Z-Z') 
\mu^{(a)}_p-(A-Z-A'+Z') \mu^{(a)}_n]/T^{(a)}}\,.
\end{equation}
Assuming NSE and considering special combinations, we can obtain 
the chemical potentials of protons ($\mu_p$) and neutrons ($\mu_n$) from the 
triton to $^4$He ratio or from the  $^3$He to $^4$He ratio, respectively, as
\begin{equation}
\mu_p^{(a)}= -19.8 {\rm MeV} +T^{(a)} \ln \left[\frac{3^{3/2}}{2^2} 
\frac{Y_\alpha}{Y_t} \right],
\end{equation}
\begin{equation}
\mu_n^{(a)}= -20.6 {\rm MeV} +T^{(a)} \ln \left[\frac{3^{3/2}}{2^2} 
\frac{Y_\alpha}{Y_h} \right].
\end{equation}
This allows us to calculate the chemical potentials separately. 
Then, considering chemical equilibrium between the different clusters, 
the total proton and neutron densities are given by the mass action law, cf. 
Eqs. (\ref{EoS:p}) and (\ref{EoS:n}). Assuming NSE where in-medium corrections 
are neglected, we find from Eq. (\ref{EoS:p0}) the total densities 
$n^{(a)}_p(T^{(a)}, \mu_p^{(a)}, \mu_n^{(a)})$ and  
 $n^{(a)}_n(T^{(a)}, \mu_p^{(a)}, \mu_n^{(a)})$ of protons and neutrons, 
respectively.

Taking into account the in-medium quasiparticle energy shifts of the nuclei, 
the relations are changed so that
\begin{equation}
\mu_p^{}= -19.818 {\rm MeV} +\Delta E_\alpha^{\rm qu}-\Delta E_t^{\rm qu} + 
T^{} \ln \left[\frac{3^{3/2}}{2^2} \frac{Y_\alpha}{Y_t} \right],
\end{equation}
\begin{equation}
\mu_n^{}= -20.582 {\rm MeV} +\Delta E_\alpha^{\rm qu}-\Delta E_h^{\rm qu} + 
T^{} \ln \left[\frac{3^{3/2}}{2^2} \frac{Y_\alpha}{Y_h} \right],
\end{equation}
where the temperature $T$ is obtained taking the medium modifications of the 
energies of nuclei into account. Now, the total proton and neutron densities 
are calculated from the EoS, Eqs. (\ref{EoS:pqu}),
which contain medium-dependent quasicluster energy shifts. 

To show the effect of these medium modifications, we start with given values 
for $T, n_p$ and $ n_n$ and calculate the cluster abundances solving 
Eqs. (\ref{EoS:pqu}),
taking the shifts into account and restricting our consideration to $A \le 4$. 
This gives us certain values for the chemical potentials $\mu_p$ and $\mu_n$.
Obviously, within a self-consistent calculation we can not only reproduce the 
input quantities $n_p$ and $n_n $ from these values of $\mu_p$ and $\mu_n$, 
but also the single ratios for different yields, in particular 
$R_{^4{\rm He},^3{\rm H}}$,  $R_{^4{\rm He},^3{\rm He}}$, and the double 
ratio $R_{\rm HHe}$. Now, we consider this as input and determine within the 
simple NSE the ACCR values $T^{(a)},\,\mu_p^{(a)}$ and $\mu_n^{(a)}$. In NSE, 
where medium shifts are neglected, we calculate 
the number densities $n^{(a)}(A,Z)$ of the nuclei, using the EoS, 
Eq. (\ref{EoS:p0}). Obviously, the single and 
double ratios given above are reproduced. However, not only the temperature 
$T^{(a)}$ will differ from the input value $T$, but also the total proton 
density $n^{(a)}_p$ and the total neutron density $n^{(a)}_n$ will deviate 
from the input values $n_p$ and $ n_n$, respectively. In Figure 2 we show the 
ratio between the ACCR baryon density $n^{(a)}_B = n^{(a)}_p + n^{(a)}_n$ 
and $n_B = n_p + n_n$ as a function of $n_B$ for various values of $T$.

\section{Discussion and Conclusions}

The assumption of NSE provides a simple means to estimate the thermodynamic 
parameters of nuclear matter at freeze-out from the observed yields of 
nuclei. This approach is applicable as long as the interaction between the 
clusters can be neglected. However, the thermometers and chemical potentials 
are no longer correctly scaled when the shifts of the binding energies due 
to the interaction with the surrounding matter become of relevance. The 
derivation of the thermodynamic parameters from the measured yields has to 
be carried out in a self-consistent manner since the binding energies, 
which determine the yields, are themselves dependent on the temperatures and 
densities.

Analyzing empirical data, the use of the ACCR method can only give a 
first approximation to the temperature and the density. Taking these first 
estimations, the shift of the binding energies of the clusters can be 
estimated. With these modified energies, the next iteration deriving the 
values of the parameters from the measured yields can be made, and a 
self-consistent solution is expected after a sufficient number of iterations. 
Alternatively one can also produce tables for yields taking the medium shifts 
into account, so that the optimal values of the parameters are obtained by 
interpolating within the table to identify the values of the parameters which 
best corresponds to the measured yields. 

Comparing the values of the parameters obtained in the full calculation, 
with inclusion of medium effects on  the yields with those deduced in 
the ACCR approach, we find that moderate deviations in the temperature 
arise for densities larger than 0.0001 fm$^{-3}$. Determination of the 
densities is more sensitive to the medium effects. 

The shift of the binding energies has been given in first order of the 
density, and higher orders terms in the density are expected to contribute 
if the density increases. Starting with baryonic densities near 
$10^{-2}$ fm$^{-3}$, the composition has to be calculated with momentum 
dependent shifts instead of the rigid shifts considered here, and then 
the temperature is found from the ratio of the mass fractions after the 
composition is calculated in a self-consistent way. One has to perform the 
full momentum integration instead of considering a rigid shift as given at 
$P=0$, when the shifts depend  on the center-of-mass momentum of the 
cluster. The results given here are applicable at densities which are 
not too high, i.e. up to 0.01 fm$^{-3}$. 

In conclusion we point out that the fragment yields from hot and dense 
nuclear matter produced in heavy ion collisions can be used to infer 
temperatures and proton/neutron densities of the early stages of the 
expanding hot matter. The assumption of thermal equilibrium can be 
only a first approach to this non-equilibrium process. To determine the 
yield of the different clusters, a simple statistical model neglecting all 
medium effects, i.e., treating it as an ideal mixture of non-interacting 
nuclei, is not applicable when the density is larger than 0.0001 fm$^{-3}$. 
Self-energy and Pauli blocking will lead to energy shifts, which have to be 
taken into account to reconstruct the thermodynamic parameters from 
measured yields. The success of the simple ACCR method to derive the values 
for the temperature can be understood from a partial compensation of the 
effect of the energy shifts so that reasonable values for the temperature 
are obtained also at relatively high densities. More care must be taken in 
inferring densities from the data. It should be mentioned that similar 
questions have to be considered when hadron production is investigated at 
the quark-gluon phase transition.   

Cross-checks can be performed to see to what extent the approach given here 
is consistent. Hitherto we considered only the yields of $d, t, h$ and 
$\alpha$, and the corresponding ratios are reflected by the temperature 
and the chemical potentials of the neutrons and protons. The determination 
of the yields of additional clusters will allow for a comparison between 
predictions and experimental data.

\begin{acknowledgments}

S. Shlomo would like to thank the Institut fur Physik of the University of 
Rostock, Rostock, Germany, for the kind hospitality. This work was supported 
by the US Department of Energy under grant No. DE-FG03-93ER40773 and the 
Robert A. Welch Foundation under grant No. A0330.

\end{acknowledgments}

\newpage

\begin{figure}
\includegraphics[width=0.85\linewidth, angle=-90, clip=true]{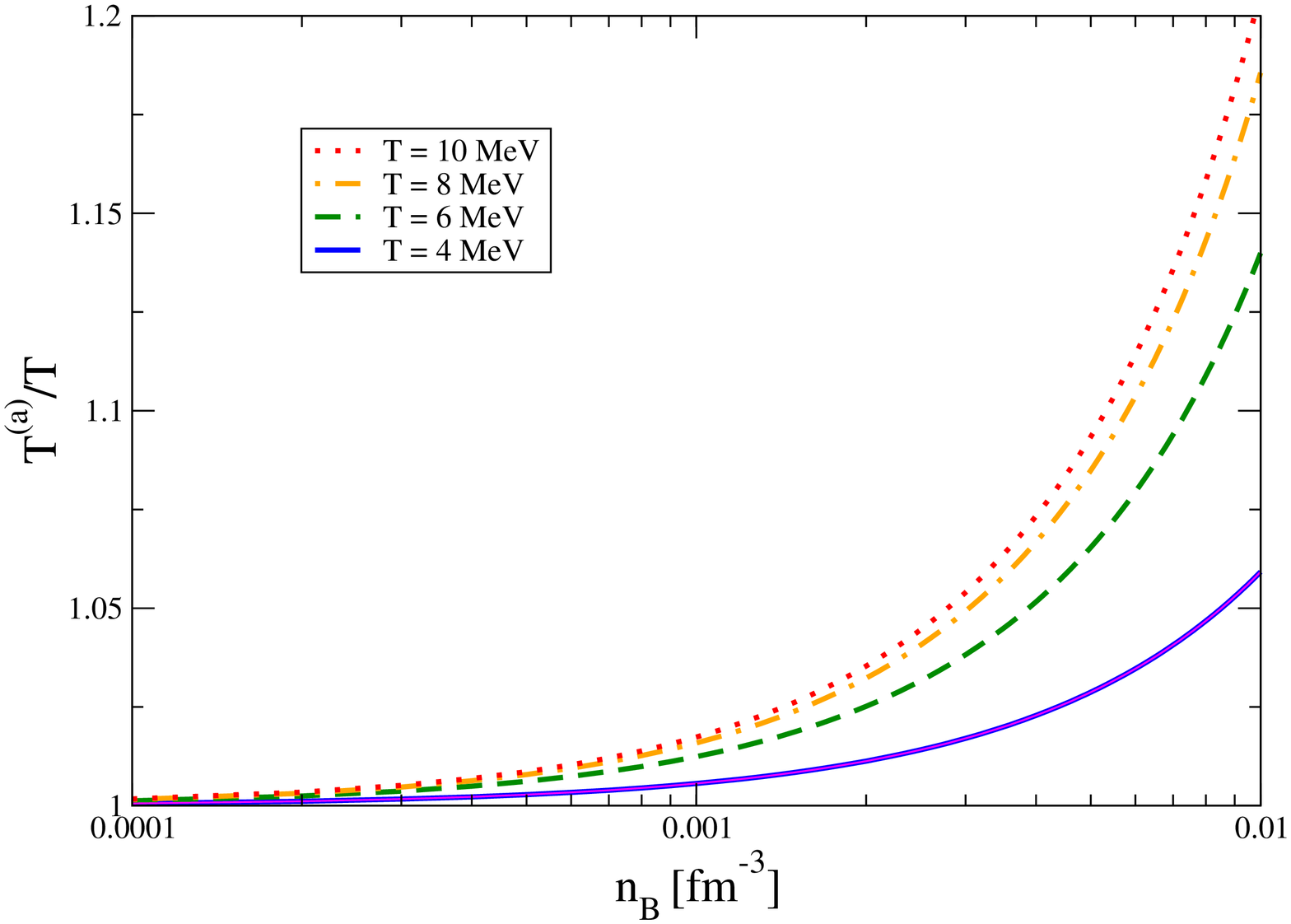}
\caption{\label{Fig. 1} (color online) The ratio between the ACCR temperature 
$T^{(a)}$ (no medium effects) and $T$ (including medium effects) as a function 
of the baryon density $n_B = n_p + n_n$ for various values of $T$.}
\end{figure}

\begin{figure}
\includegraphics[width=0.85\linewidth, angle=-90, clip=true]{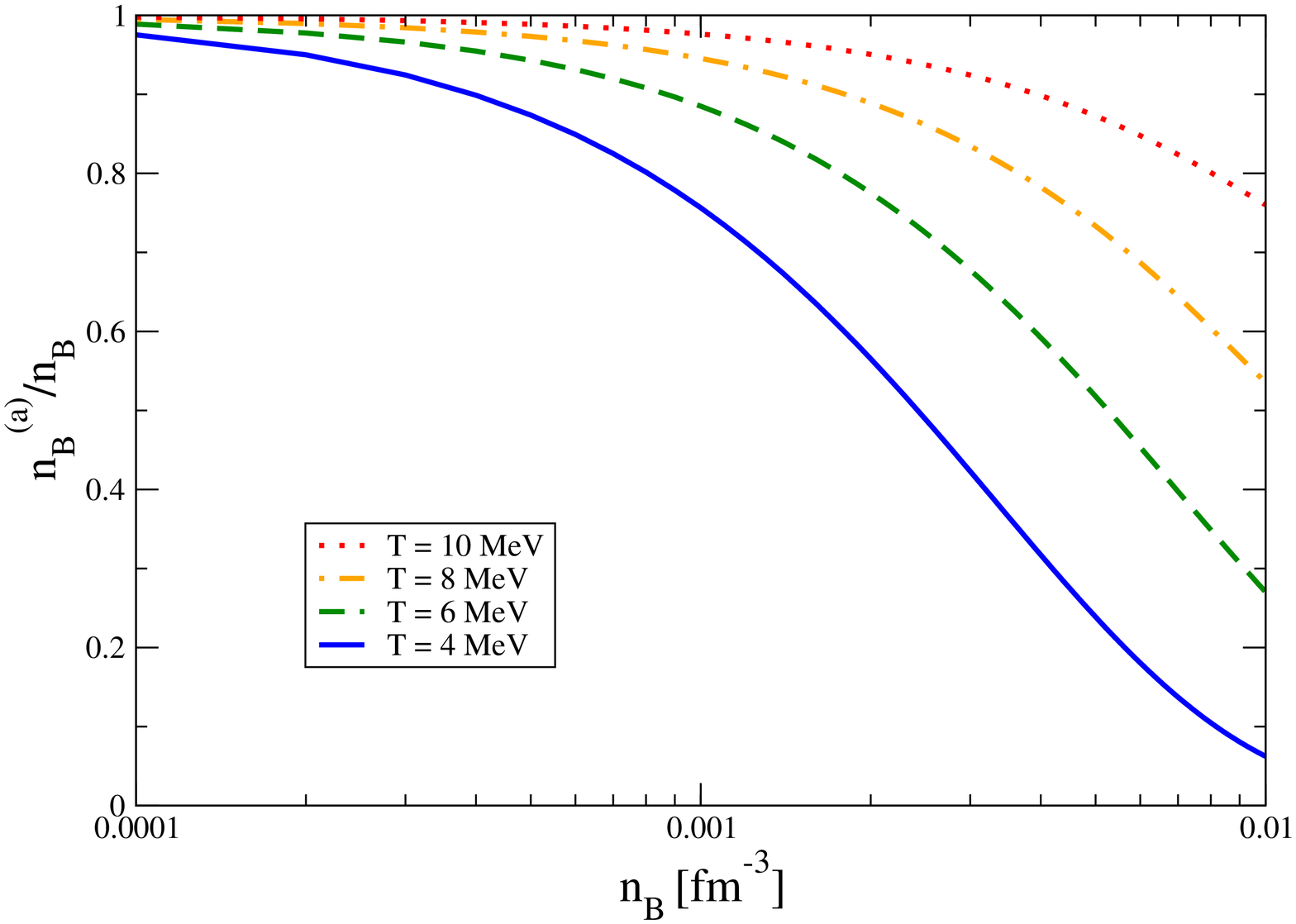}
\caption{\label{Fig. 2} (color online) Similar to figure 1, for the ratio 
$n^{(a)}_B/n_B$.}
\end{figure}


\begin{thebibliography}{20}

\bibitem{kow07} 
S. Kowalski {\it et al}, Phys. Rev. {\bf C 75}, 014601 (2007).
\bibitem{Bondorf} 
J. P. Bondorf, A. S. Botvina, A. S. Iljinov, I. N. Mishustin and K. Sneppen, 
Phys. Rep. {\bf 257}, 133 (1995).
\bibitem{Gross} 
D. H. Gross, Rep. Prog. Phys. {\bf 53}, 167 (1990).
\bibitem{albergo} 
S. Albergo, S. Costa, E. Costanzo and A. Rubbino, Nuovo Cimento {\bf A 89}, 
1 (1985).
\bibitem{Poch95} 
J. Pochodzalla {\it et al}, Phys. Rev. Lett. {\bf 75}, 1040 (1995).
\bibitem{kolom96}
A. Kolomiets {\it et al}, Phys. Rev. {\bf C 54}, R472 (1996).
\bibitem{Hauger96} 
J. A. Hauger {\it et al}, Phys. Rev. Lett. {\bf 77}, 235 (1996).
\bibitem{kolom97}
A. Kolomiets, V. M. Kolomietz and S. Shlomo, Phys. Rev. {\bf C 55}, 1376 
(1997).
\bibitem{WS34}
E. Wigner and F. Seitz, Phys. Rev. {\bf 46}, 509 (1934).
\bibitem{shlomo97} 
S. Shlomo, J. N. De and A. Kolomiets, Phys. Rev. C {\bf 55}, R2155 (1997).
\bibitem{shlomo05} 
S. Shlomo and V. M. Kolomietz, Rep. Prog. Phys. {\bf 68}, 1 (2005).
\bibitem{roe08}
G. R{\"o}pke, Phys. Rev. C, in production; arXive nucl-th/0810.4645.
\bibitem{roe83}
G. R{\"o}pke, M. Schmidt, L. M{\"u}nchow and H. Schulz, Nucl. Phys. 
{\bf A399}, 587 (1983).
\bibitem{Saha}
M. N. Saha, Phil. Mag. {\bf 40}, 472 (1920); Zeit. fur Physik {\bf 6}, 40 
(1921).
\bibitem{BraunMunzinger}
P. Braun-Munzinger, K. Redlich and J. Stachel, in {\it Quark Gluon Plasma 3}, 
Eds. R. Hwa, X.-N. Wang, (World Scientific, Singapore, 2004) P. 491 
[arXive nucl-th/0304013].
\bibitem{ChenHan}
X. Chen and P. Han, J. Phys. D: Appl. Phys. {\bf 32}, 1771 (1999).
\bibitem{hor06}
C. J. Horowitz and A. Schwenk, Nucl. Phys. {\bf A 776},55 (2006). 
\bibitem{roe05}
G. R{\"o}pke, A. Grigo, K. Sumiyoshi and Hong Shen, Part. and Nucl. Lett. 
{\bf 2}, 275 (2005). 
\bibitem{roe06}
G. R{\"o}pke, A. Grigo, K. Sumiyoshi and Hong Shen, in {\it Superdense QCD 
Matter and Compact Stars}, Ed.: D. Blaschke and A. Sedrakian 
(Springer, Dordrecht 2006), p. 75.
\bibitem{sumi08}
K. Sumiyoshi and G. R{\"o}pke, Phys. Rev. {C 77}, 055804 (2008).
\bibitem{vau72}
D. Vautherin and D. M. Brink, Phys. Lett. {\bf B 32}, 149 (1970); Phys. Rev. 
{\bf C5}, 626 (1972).
\bibitem{typ05}
S. Typel, Phys. Rev. {\bf C71}, 064301 (2005).
\bibitem{mar07}
J. Margueron, E. van Dalen and C. Fuchs, Phys. Rev. {C 76}, 034309 (2007).
\bibitem{sch90}
M. Schmidt, G. R{\"o}pke and H. Schulz, Ann. Phys. (N.Y.) {\bf 202}, 57 (1990).
\bibitem{roe84}
G. R{\"o}pke, M. Schmidt and H. Schulz, Nucl. Phys. {\bf A424}, 594 (1984).

\end{thebibliography}
\end{document}